\documentclass[paper,twocolumn,twoside]{geophysics}
\usepackage{graphicx}
\usepackage{subfigure}
\usepackage{amsfonts}
\usepackage{lipsum}
\usepackage{float}
\usepackage{amsmath}
\usepackage{nicefrac}
\usepackage{xcolor}
\usepackage{hyperref}
 \usepackage{multirow}
\usepackage{algorithm}
\usepackage{diagbox}
\usepackage[noend]{algpseudocode}
\usepackage{arydshln}
\usepackage{booktabs}

\usepackage{enumitem}

\usepackage{url}

\usepackage{hyperref}
\begin{document}

	\onecolumn

	\begin{description}[labelindent=0cm,leftmargin=3cm,rightmargin=3cm,style=multiline]

		\item[\textbf{Citation}]{M. Alfarraj and G. AlRegib, (2019), "Semisupervised sequence modeling for elastic impedance inversion," \textit{Interpretation} 7: SE237-SE249. }

		\item[\textbf{Review}]{Accepted on: 29 May 2019}

		\item[\textbf{Data and Codes}]{\href{https://github.com/olivesgatech/Elastic-Impedance-Inversion-Using-Recurrent-Neural-Networks}{[\underline{GitHub Link}]}}% If you do not have data related to this paper, you can remove the data keyword.

		\item[\textbf{Bib}] \texttt{\{@article{alfarraj2019semi,\\
  title=\{Semi-supervised Sequence Modeling for Elastic Impedance Inversion\},\\
  author=\{Alfarraj, Motaz and AlRegib, Ghassan\},\\
  journal=\{Interpretation\},\\
  volume=\{7\},\\
  number=\{3\},\\
  pages=\{SE237--SE249\},\\
  year=\{2019\},\\
  publisher={Society of Exploration Geophysicists and American Association of Petroleum Geologists\}\\}
}}

%		\item[\textbf{Copyright}]{\textcopyright 2018 IEEE. Personal use of this material is permitted. Permission from IEEE must be obtained for all other uses, in any current or future media, including reprinting/republishing this material for advertising or promotional purposes,
%			creating new collective works, for resale or redistribution to servers or lists, or reuse of any copyrighted component
%			of this work in other works. }

		\item[\textbf{Contact}]{\href{mailto:motaz@gatech.edu}{motaz@gatech.edu}  OR \href{mailto:alregib@gatech.edu}{alregib@gatech.edu}\\ \url{http://ghassanalregib.info/} \\ }
	\end{description}

	%Following command sequence was used to start the paper content from the following page and avoid numbering cover page.
	\thispagestyle{empty}
	\newpage
	\clearpage
	\setcounter{page}{1}

	%Cover page was 1 column. \twocolumn changes the page format back to double column.
	\twocolumn
\title{Semi-supervised Sequence Modeling for Elastic Impedance Inversion}

\renewcommand{\thefootnote}{\fnsymbol{footnote}}

\address{
\footnotemark[1]Center for Energy and Geo Processing (CeGP),
Georgia Institute of Technology, Atlanta, GA}
\author{Motaz Alfarraj\footnotemark[1] and Ghassan AlRegib\footnotemark[1]}

\footer{Example}
\lefthead{Alfarraj \& AlRegib}
\righthead{Semi-supervised Sequence Modeling for Elastic Impedance Inversion}

\maketitle

\begin{abstract}
Recent applications of machine learning algorithms in the seismic domain have shown great potential in different areas such as seismic inversion and interpretation. However, such algorithms rarely enforce geophysical constraints; the lack of which might lead to undesirable results. To overcome this issue, we propose a \emph{semi-supervised sequence modeling framework based on recurrent neural networks for elastic inversion from multi-angle seismic data}. Specifically, seismic traces and elastic impedance traces are modeled as time series. Then, a neural-network-based inversion model comprising convolutional and recurrent neural layers is used to invert seismic data for elastic impedance. The proposed workflow uses well-log data to guide the inversion. In addition, it utilizes seismic forward modeling to regularize the training, and to serve as a geophysical constraint for the inversion. The proposed workflow achieves an average correlation of $98\%$ between the estimated and target elastic impedance using 10 well-logs for training on a synthetic dataset.

\end{abstract}

\section{Introduction}
Seismic inversion is a process used to estimate rock properties from seismic reflection data. For example, seismic inversion can be used to infer acoustic impedance (AI) from zero-offset seismic data, which in turn is used to estimate porosity. AI is the product of P-wave velocity ($V_p$) and bulk density ($\rho$). An extension of AI for multi-angle seismic data is elastic impedance (EI) \cite[]{connolly1999elastic}. EI is a function of P-wave velocity ($V_p$), S-wave velocity ($V_s$), density ($\rho$), and incident angle ($\theta$). EI reduces to AI when $\theta=0$ \cite[]{whitcombe2002elastic}, and it utilizes information from multi-offset/angle seismic data. Thus, inverting for EI is a more powerful tool for reservoir characterization compared to AI inversion. 

The goal of inversion is to infer true model parameters ($m\in X$) through an indirect set of measurements $d \in Y$. Mathematically, the problem can be formulated as follows
\begin{equation}
    d = \mathcal{F}(m) + n,
    \label{eqn:system_setup}
\end{equation}

where $\mathcal{F}: X\rightarrow Y$ is a forward operator, $d$ is the measured data, $m$ is the model, and $n\in Y$ is a random variable that represents noise in the measurements. To estimate the model from the measured data, one needs to solve an inverse problem. The solution depends on the nature of the forward model and observed data. In the case of seismic inversion, and due to the non-linearity and heterogeneity of the subsurface, the inverse problem is highly ill-posed. In order to find a stable solution to an ill-posed problem, the problem is often regularized by imposing constraints on the solution space, or by incorporating prior knowledge about the model. 
A classical approach to solve inverse problems is to set up the problem as a Bayesian inference problem, and improve prior knowledge by optimizing for a cost function based on the data likelihood,
\begin{equation}
    \hat{m} = \underset{m\in X}{\text{argmin}} \left[ \mathcal{H}\left(\mathcal{F}(m),d\right)+ \lambda \mathcal{C}(m)\right],
    \label{eqn:classical_inversion}
\end{equation} 

where $\hat{m}$ is the estimated model, $\mathcal{H}:Y\times Y \rightarrow \mathbb{R}$ is an affine transform of the data likelihood, $\mathcal{C}:X \rightarrow \mathbb{R}$ is a regularization function that incorporates prior knowledge, and $\lambda\in \mathbb{R}$ is regularization parameter that controls the influence of the regularization function. 

The solution of equation \ref{eqn:classical_inversion} in seismic inversion can be sought in a stochastic or a deterministic fashion through an optimization routine. In stochastic inversion, the outcome is a posterior probability density function; or multiple valid solutions to account for uncertainty in the data. On the other hand, deterministic inversion produces an estimate ($\hat{m}$) that maximizes the posterior probability density function. The literature of seismic inversion (both deterministic and stochastic) is rich in various methods to formulate, regularize and solve the problem (e.g., \cite[]{duijndam1988bayesian_1, doyen1988porosity, duijndam1988bayesian_2, ulrych2001bayes, buland2003bayesian, tarantola2005inverse,doyen2007seismic, bosch2010seismic, gholami2015nonlinear, azevedo2017geostatistical}).

Recently, there have been several successful applications of machine learning and deep learning methods to solving inverse problems \cite[]{lucas2018using}. Moreover, machine learning and deep learning methods have been utilized in the seismic domain for different tasks such as inversion and interpretation \cite[]{alregib2018subsurface}. For example, seismic inversion has been attempted using supervised-learning algorithms such as support vector regression (SVR) \cite[]{al2012support}, artificial neural networks \cite[]{roth1994neural,araya2018deep}, committee models \cite[]{gholami2017estimation}, convolutional neural networks (CNNs) \cite[]{das2018convolutional}, and many other methods \cite[]{chaki2015novel,yuan2013spectral, gholami2017estimation,chaki2017diffusion,mosser2018rapid, chaki2018well}. More recently, a sequence-modeling-based machine learning workflow was used to estimate petrophysical properties from seismic data \cite[]{alfarraj2018petrophysical}, which showed that recurrent neural networks are superior to feed-forward neural networks in capturing the temporal dynamics of seismic traces.

In general, machine learning algorithms can be used to invert seismic data by learning a non-linear mapping parameterized by $\Theta\in Z \subseteq \mathbb{R}^n$, i.e., $\mathcal{F}_{\Theta}^\dagger: Y\rightarrow X$ from a set of examples (known as a training dataset) such that:
\begin{equation}
    \mathcal{F}_{\Theta}^\dagger(d) \approx m.
    \label{eqn:machine_learning_inversion}
\end{equation}

A key difference between classical inversion methods and learning-based methods is the outcome. In classical inversion, the outcome is a set of model parameters (deterministic) or a posterior probability density function (stochastic). On the other hand, learning methods produce a mapping from the measurements domain (seismic) to model parameters domain (rock property). Another key difference between the two approaches is their sensitivity to the initialization of the optimization routine. In classical inversion, the initial guess (posed as a prior model) plays an important role in the convergence of the method, and in the final solution. On the other hand, learning-based methods are often randomly initialized, and prior knowledge is integrated into the objective function and is inferred by the learning algorithm from the training data. Thus, learning-based inversion is less sensitive to the initial guess. On the other hand, learning-based inversion requires a high-quality training dataset in order to generate reliable results. Nevertheless, there have been efforts to overcome this shortcoming of neural networks and enable them to learn from noisy or unreliable data \cite[]{natarajan2013learning}.

There are many challenges, however, that might prevent machine learning algorithms from finding a proper mapping that can be generalized for an entire survey area. One of the challenges is the lack of data from a given survey area on which an algorithm can be trained. For this reason, such algorithms must have a limited number of learnable parameters and a good regularization method in order to prevent over-fitting and to be able to generalize beyond the training data. Moreover, applications of machine learning on seismic data do not often utilize or enforce a physical model that can be used to check the validity of their outputs. In other words, there is a high dependence on machine learning algorithms to understand the inherent properties of the target outputs without explicitly specifying a physical model. Such dependence can lead to undesirable or incorrect results, especially when data is limited. 

In this work, we propose a semi-supervised learning workflow for elastic impedance (EI) inversion from multi-angle seismic data using sequence modeling through a combination of recurrent and convolutional neural networks. The proposed workflow learns a non-linear inverse mapping from a training set consisting of well-log data and their corresponding seismic data. In addition, the learning is guided and regularized by a forward model as often incorporated in classical inversion theory. The rest of this paper is organized as follows. First, we discuss the formulation of learning-based seismic inversion. Then, we introduce recurrent neural networks as one of the main components in the proposed workflow. Next, we discuss the technical details of the proposed workflow. Then, we present a case study to validate the proposed framework on the Marmousi 2 model \cite[]{martin2006marmousi2} by inverting for EI using seismic data and 10 well-logs only.

\section{Problem Formulation}
Using neural networks, one can learn the parameter $\Theta$ (in equation \ref{eqn:machine_learning_inversion}) in either a supervised manner or an unsupervised manner \cite[]{adler2017solving}. In supervised learning, the machine learning algorithm is given a set of measurement-model pairs $(d,m)$ (e.g., seismic traces and their corresponding property traces from well-logs). Then, the algorithm learns the inverse mapping by minimizing the following loss function
\begin{equation}
    L_1(\Theta):=\sum_{\substack{i\\m_i\in \mathcal{S}}} \mathcal{D}\left(m_i, \mathcal{F}_{\Theta}^\dagger(d_i)\right)
    \label{eqn:supervised}
\end{equation}

where $\mathcal{S}$ is the set of available property traces from well-logs, $m_i$ is the $i^\text{th}$ trace in $\mathcal{S}$, $d_i$ is the seismic traces corresponding to $m_i$, and $\mathcal{D}$ is a distance measure that compares the estimated rock property to the true property. Namely, supervised machine learning algorithms seek a solution that minimizes the inversion error over the given measurement-model pairs. Note that equation \ref{eqn:supervised} is computed only over a subset of all traces in the survey. This subset includes the traces for which a corresponding property trace is available from well-logs. 

Alternatively, the parameters $\Theta$ can be sought in an unsupervised-learning scheme where the learning algorithm is given a set of measurements ($d$) and a forward model $\mathcal{F}$. The algorithm then learns by minimizing the following loss function,
\begin{equation}
    L_2(\Theta):=\sum_{i}\mathcal{D}\left(\mathcal{F}\left(\mathcal{F}_{\Theta}^\dagger(d_i)\right),d_i\right), 
    \label{eqn:unsupervised}
\end{equation}
 
which is computed over all seismic traces in the survey. The loss in equation \ref{eqn:unsupervised} is known as data misfit. It measures the distance between the input seismic traces and the synthesized seismograms from the estimated property traces using the forward model. 

Although supervised methods have been shown to be superior to unsupervised ones in various learning tasks (e.g., image segmentation and object recognition), they often need to be trained on a large number of training examples. In the case of the seismic inversion, the labels (i.e., property traces) are scarce since they come from well-logs. Unsupervised methods, on the other hand, do not require labeled data, and thus can be trained on all available seismic data only. However, unsupervised learning does not integrate well-logs (direct model measurements) in the learning.

In this work, we propose a semi-supervised learning workflow for seismic inversion that integrates both well-log data in addition to data misfit in learning the inverse model. Formally, the loss function of the proposed semi-supervised workflow is written as 
\begin{equation}
    L(\Theta):= \alpha\cdot \underset{\text{property loss}}{\underbrace{L_1(\Theta)}} + \beta \cdot \underset{\text{seismic loss}}{\underbrace{L_2(\Theta)}}
    \label{eqn:semi-supervised}
\end{equation}

where $\alpha, \beta \in \mathbb{R}$ are tuning parameters that govern the influence of each of the property loss and seismic loss, respectively. For example, if the input seismic data is noisy, or well-log data is corrupted, the values of $\alpha$ and $\beta$ can be used to limit the role of the corrupted data in the learning process. The property loss is computed over the traces for which we have access to model measurements (rock properties) form well-logs. The seismic loss , on the other hand, is computed over all traces in the survey. 

The goal of this work is to invert for elastic impedance (EI) using multi-offset data using semi-supervised sequence modeling. Hence, ($d$) in the equation \ref{eqn:semi-supervised} is the set of all multi-angle traces in the seismic survey, and $m$ is the set of available EI traces from well-logs. Naturally, the size of $m$ is small compared to $d$. Furthermore, the vertical resolution of EI traces is finer than that of the seismic traces . There are two common ways to model traces in a learning paradigm. The first method is to treat each data point in a well-log (in the $z$-direction) as an independent sample and try to invert for a given rock property from the corresponding multi-angle seismic data sample. This method fails to capture the long-term temporal dynamics of seismic traces; that is the dependency between a data point at a given depth and the data points before it and after it. An alternative approach is to treat each trace as a single training sample to incorporate the temporal dependency. However, this approach severely limits the amount of data from which the algorithm can learn since each trace is treated as a single training sample. With a limited amount of training data, common machine learning algorithms might fail to generalize beyond the training data because of their high data requirements. Such a requirement might be difficult to meet in practical settings where the number of well-logs is limited. In order to remedy this, the proposed workflow utilizes sequence modeling to model traces as sequential data via recurrent neural networks.

\section{Recurrent Neural Networks}
Despite the success of feed-forward machine learning methods, including convolutional networks, multilayer perceptrons, and support vector machines for various learning tasks, they have their limitations. Feed-forward methods have an underlying assumption that data points are independent, which makes them fail in modeling sequentially dependent data such as videos, speech signals, and seismic traces. 

Recurrent neural networks (RNN), on the other hand, are a class of artificial neural networks that are designed to capture temporal dynamics of sequential data. Unlike feed-forward neural networks, RNNs have a hidden state variable that can be passed between sequence samples which allows them to capture long temporal dependencies in sequential data. RNNs have been utilized to solve many problems in language modeling and natural language processing (NLP)\cite[]{mikolov2010recurrent}, speech and audio processing \cite[]{graves2013speech}, video processing \cite[]{ma2017ts}, petrophysical property estimation \cite[]{alfarraj2018petrophysical}, detection of natural earthquakes \cite[]{wiszniowski2014application}, and stacking velocity estimation \cite[]{biswas2018stacking}.

A single layer feed-forward neural network produces an output $\mathbf{y}_i$ which is a weighted sum of input features $\mathbf{x}_i$ followed by an activation function (a non-linearity) like the sigmoid or hyperbolic tangent functions, i.e. $\mathbf{y}_i = \sigma \left(\mathbf{W}\mathbf{x}_i+\mathbf{b}\right)$, where $\mathbf{x}_i$ and $\mathbf{y}_i$ are the input and output feature vectors of the $i^{th}$ sample, respectively, $\sigma(\cdot)$ is the activation function, $\mathbf{W}$ and $\mathbf{b}$ are the learnable weight matrix and bias vector, respectively. The same equation is applied to all data samples independently to produce corresponding outputs. 

\begin{figure}[ht]
    \centering
    \includegraphics[width=0.8\linewidth]{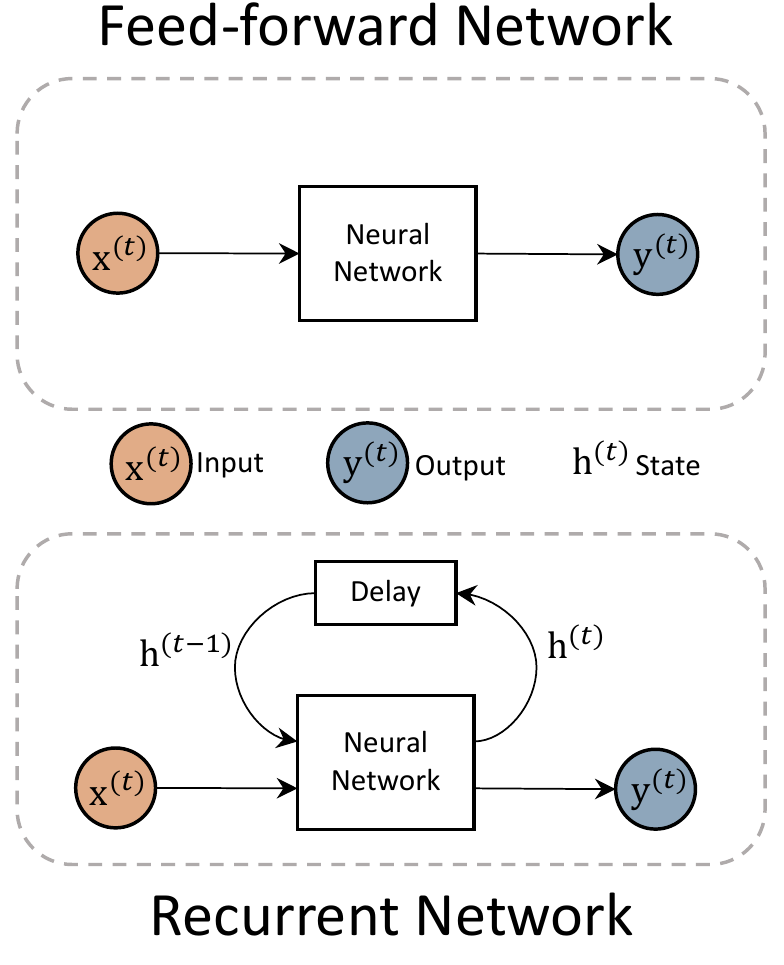}
    \caption{An illustration of feed-forward and recurrent networks.}
    \label{fig:RNN_vs_NN}
\end{figure}

In addition to the affine transformation and the activation function, RNNs introduce a hidden state variable that is computed using the current input and the hidden state variable from the previous step, 
\begin{equation}
    \begin{aligned}
    \mathbf{h}_i^{(t)} &= \sigma\left(\mathbf{W}_{xh}\mathbf{x}_i^{(t)}+\mathbf{W}_{hh}\mathbf{h}_i^{(t-1)} + \mathbf{b}_h\right), \\
    \mathbf{y}_i^{(t)} &= \sigma\left(\mathbf{W}_{hy}\mathbf{h}_i^{(t)} + \mathbf{b}_y\right) 
    \end{aligned}
    \label{eqn:RNN}
\end{equation}

where $\mathbf{x}_i^{(t)}$, $\mathbf{y}_i^{(t)}$ and $\mathbf{h}_i^{(t)}$, are the input, output, and state vectors at time step $t$, respectively, $\mathbf{W}$'s and  $\mathbf{b}$'s are network weights and bias vectors respectively. For time $t=0$, the hidden state variable is set to $\mathbf{h}^{(0)} = \mathbf{0}$. Figure \ref{fig:RNN_vs_NN} shows a side-by-side comparison between a feed-forward unit and a recurrent unit. 

When RNNs were proposed in the 1980s, they were difficult to train because of the introduced dependency between data samples that made the gradients more difficult to compute. The problem was later solved using backpropagation through time (BPTT) algorithms \cite[]{werbos1990backpropagation}, which turns gradients into long products of terms using the chain rule. Theoretically, RNNs are supposed to learn long-term dependencies from their hidden state variable. However, even with BPTT, RNNs fail to learn long-term dependencies mainly because the gradients tend to vanish for long sequences when backpropagated through time.

New RNN architectures have been proposed to overcome the issue of vanishing gradients using gated units. Examples of such architectures are Long Short-Term Memory (LSTM) \cite[]{hochreiter1997lstm} and the recently proposed Gated Recurrent Units (GRU) \cite[]{cho2014properties}. Such architectures have been shown to capture long-term dependency and perform well for various tasks such as machine translation and speech recognition. In this work, we utilize GRUs in the proposed inversion workflow.

\begin{figure}[ht]
    \centering
    \includegraphics[width=0.9\linewidth]{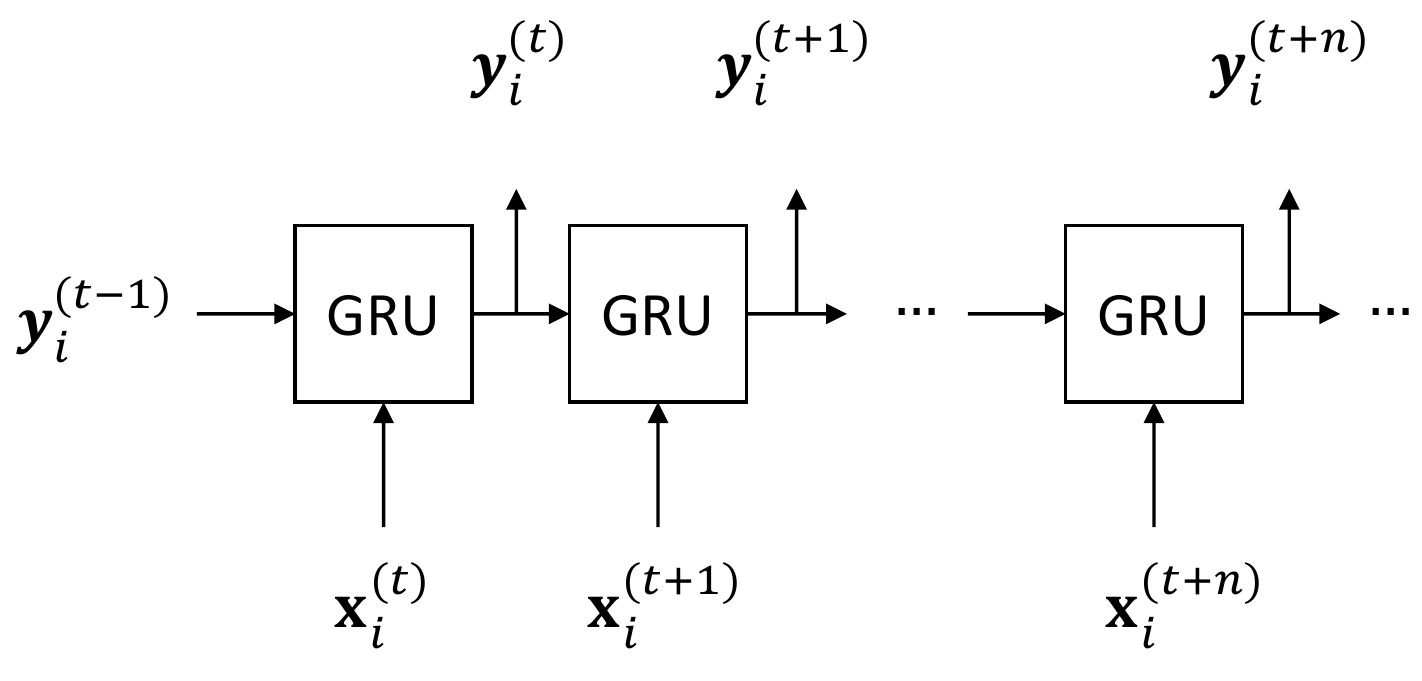}
    \caption{Gated Recurrent Unit (GRU) unfolded through time.}
    \label{fig:GRU}
\end{figure}

GRUs supplement the simple RNN described above by incorporating reset-gate and update-gate variables which are internal states that are used to evaluate the long-term dependency and keep information from previous times only if they are needed. The forward step through a GRU is given by the following equations, 
\begin{equation}
    \begin{aligned}
        \mathbf{u}_i^{(t)} &= \text{sigmoid}\left(\mathbf{W}_{xu}\mathbf{x}_i^{(t)}+\mathbf{W}_{yu}\mathbf{y}_i^{(t-1)} + \mathbf{b}_u\right) \\
        \mathbf{r}_i^{(t)} &= \text{sigmoid}\left(\mathbf{W}_{xr}\mathbf{x}_i^{(t)} + \mathbf{W}_{yr}\mathbf{y}_i^{(t-1)} + \mathbf{b}_r\right) \\
        \hat{\mathbf{y}}_i^{(t)} &=  \tanh\left(\mathbf{W}_{x\hat{y}}\mathbf{x}_i^{(t)} + \mathbf{b}_{\hat{y}_1}+  r_i^{(t)}\circ\left(\mathbf{W}_{h\hat{y}}\mathbf{y}_i^{(t-1)} +  \mathbf{b}_{\hat{y}_2}\right)\right)\\
        \mathbf{y}_i^{(t)} &= (1-\mathbf{u}^{(t)})\circ \mathbf{y}_i^{(t-1)} + \mathbf{u}^{(t)}\circ \hat{\mathbf{y}}^{(t)}
    \end{aligned}
    \label{eqn:GRU}
\end{equation}

where $\mathbf{u}_i^{(t)}$ and $\mathbf{r}_i^{(t)}$ are the update-gate and reset-gate vectors, respectively, $\hat{\mathbf{y}}_i^{(t)}$ is the candidate output for the current time step, $\mathbf{W}$'s and $\mathbf{b}$'s are the learnable parameters, and the operator $\circ$ is the element-wise product. Note that the GRU introduces two new state variables, update-gate $\mathbf{u}$ and reset-gate $\mathbf{r}$, which control the flow of information from one time step to another, and thus they are able to capture long-term dependency. The output of the GRU at the current time step ($\mathbf{y}_i^{(t)}$) is a weighted sum of the candidate output for the current time step and the output from the previous step. Figure \ref{fig:GRU} shows a GRU network unfolded through time. Note that all GRUs in an unfolded network share the same $\mathbf{W}$ and $\mathbf{u}$ parameters.

\section{Methodology}
Similar to all deep learning methods, RNNs require tremendous amounts of data to train. Given that well-log data is limited in a given survey area, the number of training samples is limited. With such limitation, a combination of regularization techniques must be used to train a learning-based model properly and ensure it generalizes beyond the training dataset \cite[]{alfarraj2018petrophysical}. In addition, the data shortage limits the number of the layers (and hence parameters) that can be used in learning-based models. Therefore, using deeper networks to capture the highly non-linear inverse mapping from seismic data to EI might not be feasible using supervised learning. 

In this work, we utilize a seismic forward model as another form of supervision in addition to well-log data. Although forward modeling is an essential block in classical seismic inversion approaches, it has not been integrated into learning-based seismic inversion workflows. 

Using a forward model in a learning-based inversion has two main advantages. First, it allows the incorporation of geophysics into a machine learning paradigm to ensure that the outputs of the networks are obeying the physical laws. In addition, it allows the algorithm to learn from all traces in the seismic survey without explicitly providing an EI trace (label) for each seismic trace.

\subsection{Proposed workflow}
\begin{figure}[h]
    \centering
    \includegraphics[width=\linewidth]{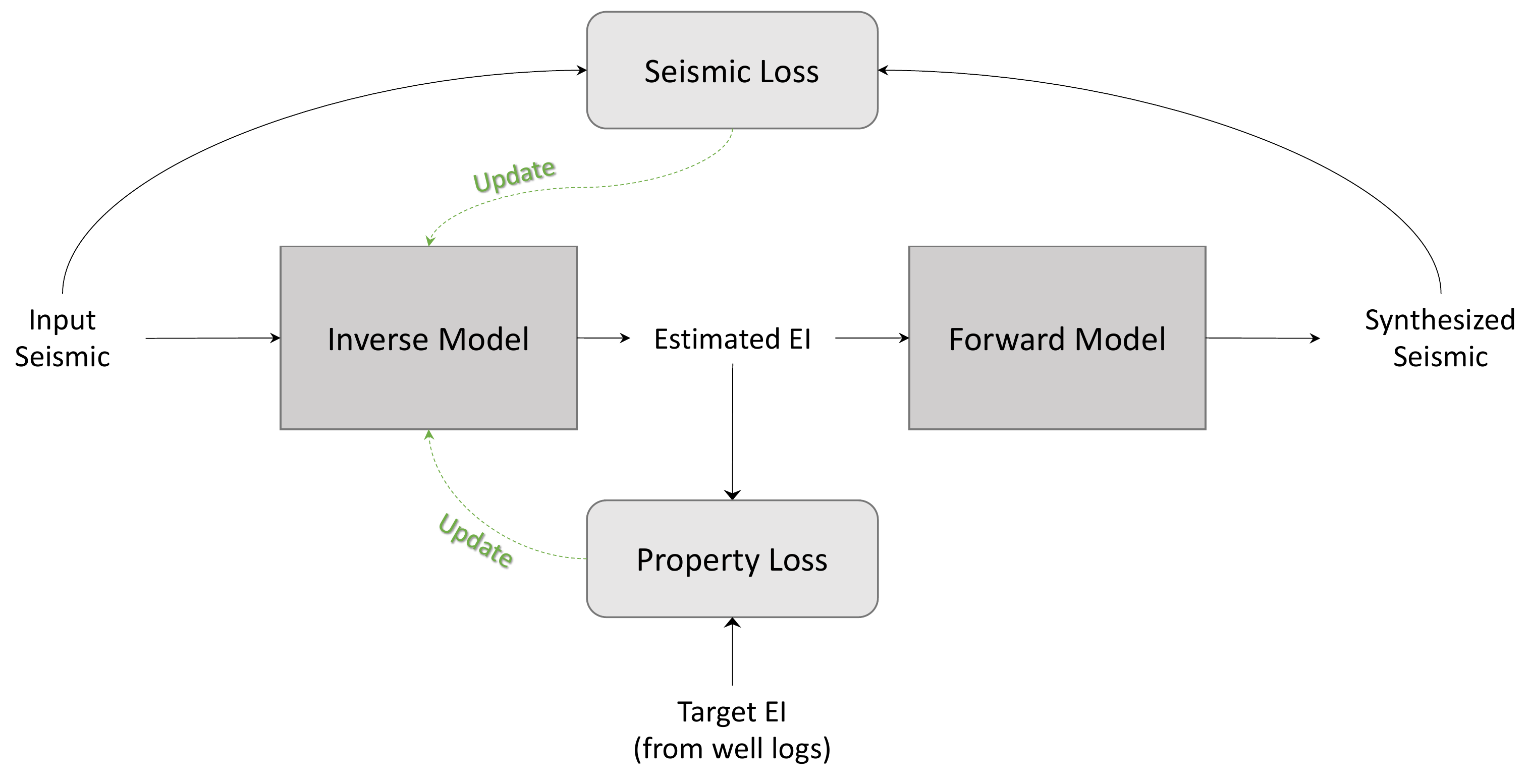}
    \caption{The proposed semi-supervised inversion workflow.}
    \label{fig:workflow}
\end{figure}

The inverse model ($\mathcal{F}_\Theta^\dagger$) can, in principle, be trained using well-log data and their corresponding seismic data only. However, as we have discussed earlier, a deep inverse model requires a large dataset of labeled data to train properly which is not possible in a practical setting where the number of well-logs is limited. Integrating a forward model, as commonly done in classical inversion workflows, allows the inverse model to learn from the seismic data without requiring their corresponding property traces, in addition to learning from the few available property traces from well-logs. The overall proposed inversion workflow is shown in Figure \ref{fig:workflow}. 

The workflow in Figure \ref{fig:workflow} consists of two main modules: the inverse model ($\mathcal{F}^\dagger_\Theta$) with learnable parameters, and a forward model ($\mathcal{F}$) with no learnable parameters. The proposed workflow takes multi-angle seismic traces as inputs, and outputs the best estimate of the corresponding EI. Then, the forward model is used to synthesize multi-offset seismograms from the estimated EI. The error (data misfit) is computed between the synthesized seismogram and the input seismic traces using the \textit{seismic loss}. This process is done for all traces in the survey. Furthermore, \textit{property loss} is computed between estimated and true EI on traces for which we have a true EI from well-logs. The parameters of the inverse model are adjusted by combining both losses as in equation \ref{eqn:semi-supervised} using a gradient-descent optimization. 

In this work, we chose the distance measure ($\mathcal{D}$) as the Mean Squared Error (MSE) and $\alpha=\beta=1$. Hence, equation \ref{eqn:semi-supervised} reduces to: 
\begin{equation}
L(\Theta) =  \frac{1}{N_p}\sum_{\substack{i\\m_i\in\mathcal{S}}}\lVert m_i - \mathcal{F}^\dagger_\Theta(d_i)\rVert_2^2 + \frac{1}{N_s}\sum_{i}\lVert d_i - \mathcal{F}(\mathcal{F}^\dagger_\Theta(d_i)) \rVert_2^2
\label{eqn:loss}  
\end{equation}

where $N_\text{s}$ is the total number of seismic traces in the survey, and $N_p = |\mathcal{S}| $ is the number of available well-logs. In seismic surveys, $N_p \ll N_s$, therefore, \emph{seismic loss} is computed over many more traces than the property loss. On the other hand, \emph{property loss} has access to direct model parameters (well-log data). These factors make the two losses work in collaboration to learn a stable and accurate inverse model. 

It is important to note that the choice of the forward model is critical in the proposed workflow for two reasons. First, the forward model must be able to synthesize at a speed comparable to the speed at which the inverse model processes data. Since deep learning models, in general, are capable of processing large amounts of data in a very short time with GPU technology, the forward model must be fast. Second, the proposed inverse model, as all other deep learning models, adjusts its parameters according to the gradients with respect to a defined loss function, therefore, the forward model must be differentiable in order to compute gradients with respect to the seismic loss. Therefore, in this work, we chose a convolutional forward model due to its simplicity and efficiency to reduce computation time. Other choices of the forward model are possible as long as they satisfy the two conditions stated above. 

\subsection{Inverse Model}
\begin{figure*}
    \centering
    \includegraphics[width=\linewidth]{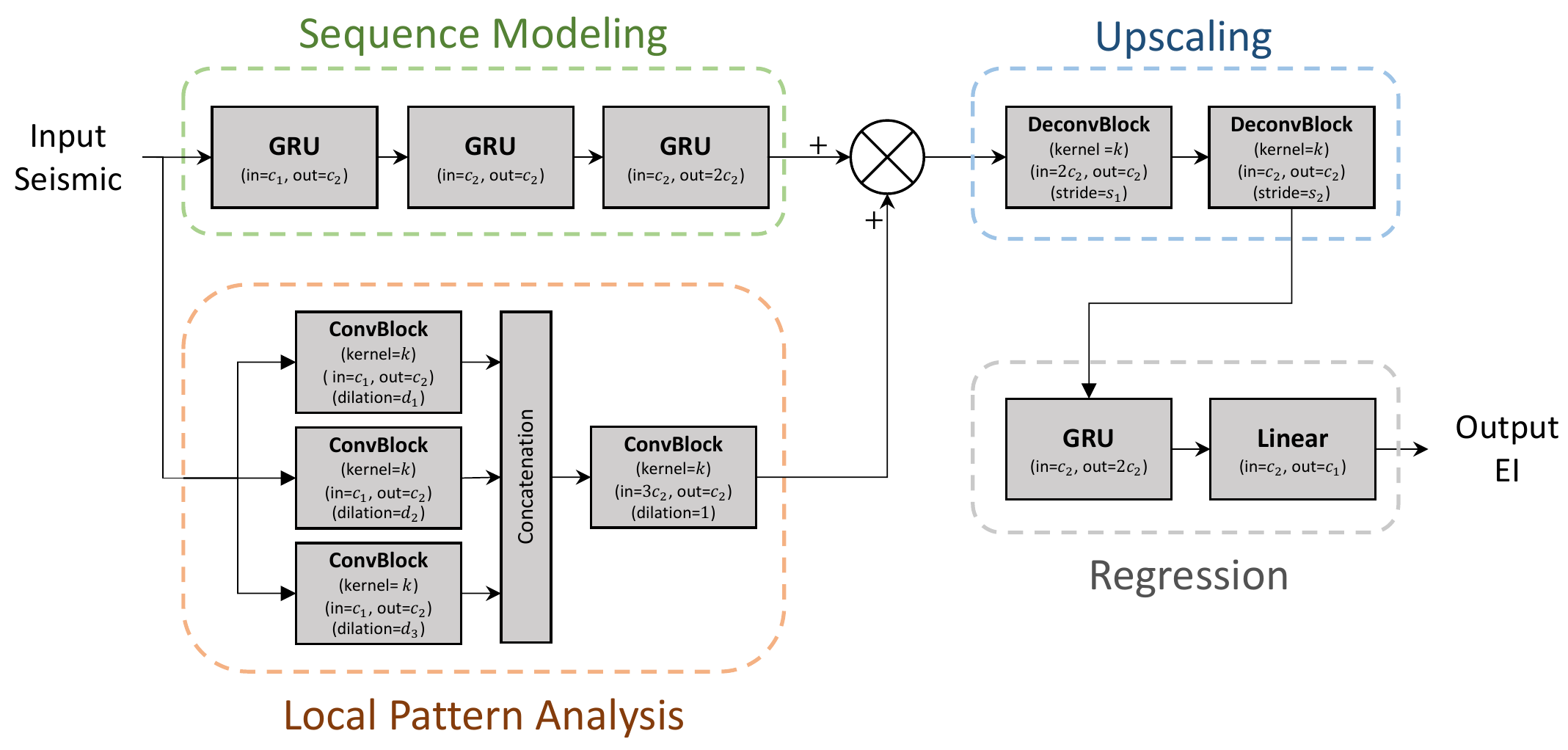}
   \caption{The inverse model in the proposed workflow with generic hyperparameters. The hyperparameters are chosen based on the data. The number of input and output features are denoted by \textsc{in} and \textsc{out}, respectively}
    \label{fig:inverse}
\end{figure*}

The proposed inverse model of the proposed workflow (shown in Figure \ref{fig:inverse}) consists of four main submodules. These submodules are labeled as \emph{sequence modeling}, \emph{local pattern analysis}, \emph{upscaling}, and \emph{regression}. Each of the four submodules performs a different task in the overall inverse model.

The \emph{sequence modeling} submodule models temporal dynamics of seismic traces and produces features that best represent the low-frequency content of EI. The \emph{local pattern analysis} submodule extracts local attributes from seismic traces that best model high-frequency trends of EI trace. The \emph{upscaling} submodule takes the sum of the features produced by the previous modules and upscales them vertically. This module is added based on the assumption that seismic data are sampled (vertically) at a lower resolution than that of well-log data. Finally, the \emph{regression} submodule maps the upscaled outputs from features domain to target domain (i.e., EI).  The details of each these submodule are discussed next.  

\subsubsection{Sequence Modeling}
The \emph{sequence modeling} submodule consists of a series of bidirectional Gated Recurrent Units (GRU). Each bidirectional GRU computes a state variable from future and past predictions and is equivalent to 2 GRUs where one models the trace from shallow to deeper layers, and the other models the reverse trace. Assuming each multi-angle seismic traces have $c_1$ channels (one channel for each incident angle), the First GRU takes these $c_1$ channels as input features and computes $c_2$ temporal features based on the temporal variations of the processed traces. The next two GRUs perform a similar task on the outputs of their respective preceding GRU. The series of the three GRUs is equivalent to a 3-layer deep GRU. Deeper networks are able to model complex input-output relationships that shallow networks might not capture. Moreover, deep GRUs generally produce smooth outputs. Hence, the output of the \emph{sequence modeling} submodule is considered as the low-frequency trend of EI. 

\subsubsection{Local pattern analysis}
\begin{figure}[ht!]
    \centering
    \includegraphics[width=0.8\linewidth]{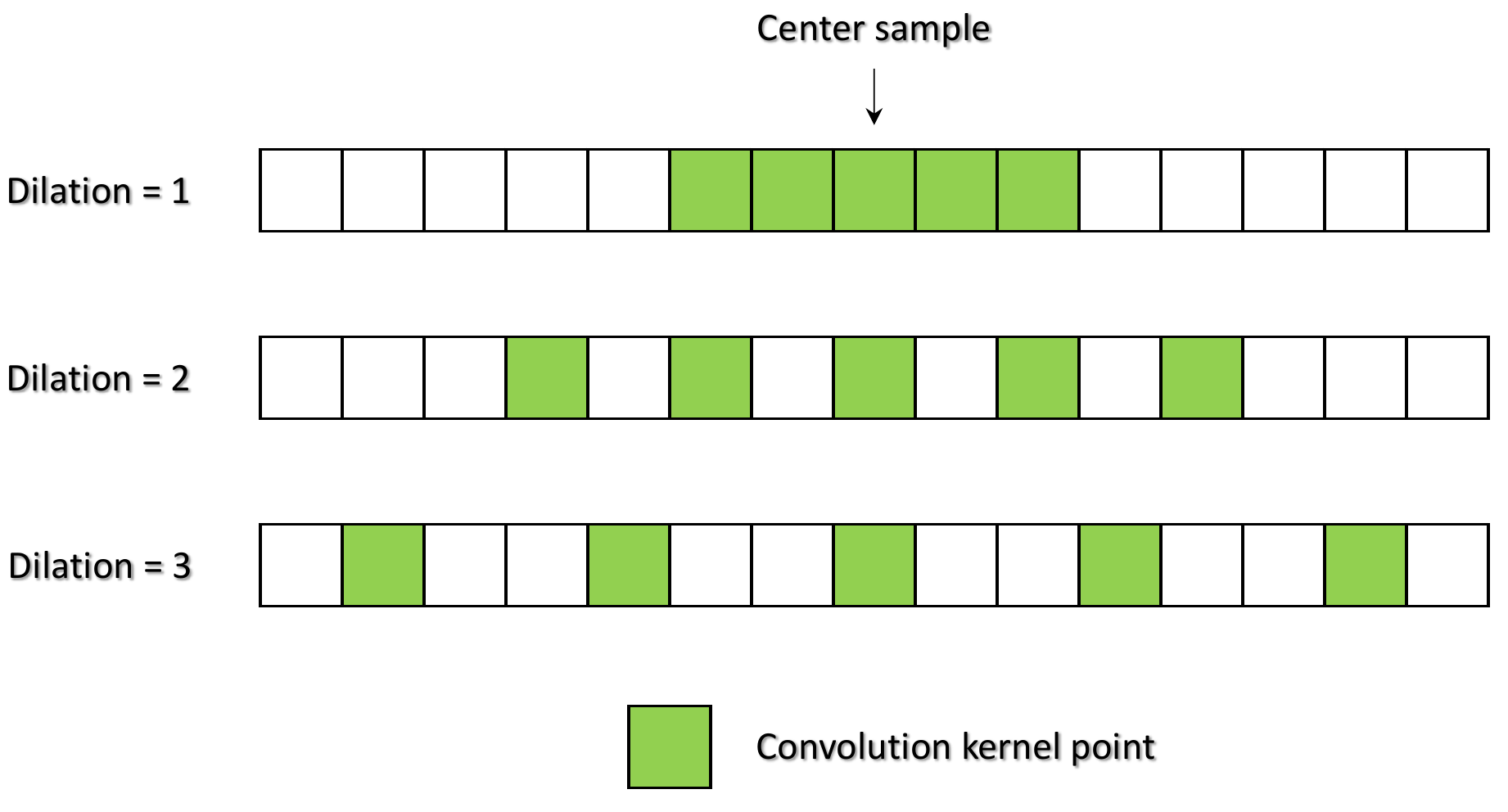}
    \caption{An illustration of dilated convolution for multiscale feature extraction (kernel size $=5$, dilation factors $\text{dilation} =1,2$ and $3$).}
    \label{fig:dilation}
\end{figure}

Another submodule of the inverse model is the \emph{local pattern analysis} submodule which consists of a set of 1-dimensional convolutional blocks with different dilation factors in parallel. The output features of each of the parallel convolutional blocks are then combined using another convolutional block. Dilation refers to the spacing between convolution kernel points in the convolutional layers \cite[]{yu2015multi}. Multiple dilation factors of the kernel extract multiscale features by incorporating information from trace samples that are direct neighbors to a reference sample (i.e., the center sample), in addition to the samples that are further from it. An illustration of dilated convolution is shown in Figure \ref{fig:dilation} for a convolution kernel of size $5$ and dilation factors $\text{dilation} = 1,2$ and $3$.

A convolutional block (\textit{ConvBlock}) consists of a convolutional layer followed by group normalization \cite[]{wu2018group} and an activation function. Group normalization scales divides the output of the convolutional layers into groups, and normalizes each group using a learned mean and standard deviation. They have been shown to reduce covariant shift in the learned features and speed up the learning. In addition, activation functions are one of the building blocks of any neural network. They are a source of non-linearity that allow the neural networks to approximate highly non-linear functions. In this work, we chose the hyperbolic tangent function as the activation function. 

Convolutional layers operate on small windows of the input trace due to their small kernel sizes. Therefore, they capture high-frequency content. Since convolutional layers do not have a state variable like recurrent layers to serve as a memory, they do poorly in estimating the low-frequency content. The outputs of the \emph{local pattern analysis} submodule are of very similar dimensions of those of the \emph{sequence modeling} submodule. Hence, the outputs of the two modules are added to obtain a full-band frequency content.  

\subsubsection{upscaling}
Seismic data is sampled at a lower rate than that of well-logs data. The role of the \emph{upscaling} submodule is to compensate for this resolution mismatch. This submodule consists of two Deconvolutional Blocks with different kernel stride. The stride controls the factor by which the inputs are upscaled. A stride of ($s=2$) deconvolutional block produces an output that has twice the number of the input samples (vertically). 

Deconvolutional layers (also known as transposed convolutional or fractionally-strided convolutional layers) are upscaling modules with learnable kernel parameters unlike classical interpolation methods with fixed kernel parameters (e.g., linear interpolation). They learn kernel parameters from both feature and local spatial/temporal domain. They have been used for various applications like semantic segmentation and seismic structure labeling \cite[]{noh2015learning,alaudah2018learning}. 

Deconvolutional blocks in Figure \ref{fig:inverse} have a similar structure as the convolutional blocks introduced earlier. They are a series of deconvolutional layer followed by a group normalization module and an activation function. 

\subsubsection{Regression}
The final submodule in the inverse model is the \emph{regression} submodule which consists of a GRU followed by a linear mapping layer (fully-connected layer). Its role is to regress the extracted features from the other modules to the target domain (EI domain). The GRU in this module is a simple 1-layer GRU that augments the upscaled outputs using global temporal features. Finally, a linear affine transformation layer (fully-connected layer) takes the output features from the GRU and maps them to the same number of features in the target domain, which is, in this case, the number of incident angles in the target EI trace. 

\subsection{Forward Model}
Forward modeling is the process of synthesizing seismograms from elastic properties of the earth (i.e., P-wave velocity, S-wave velocity, and density) or from a function of the elastic properties such as EI. In this work, and since we are inverting for EI, we used a forward model that takes EI as an input, and outputs a corresponding seismogram. EI was proposed by \cite{connolly1999elastic} and later normalized by \cite{whitcombe2002elastic}. It is based on the Aki-Richards approximation for Zoeppritz equations \cite[]{aki1980quantitative}. The Aki-Richards approximation incorporates amplitude variations with offset/angle (AVO/AVA) based on the changes of elastic properties and the incident angle. 

The forward model adopted in this work uses Connolly's formulation to compute the reflection coefficients from $EI$ for different incident angles $\theta$ as follows:

\begin{equation}
    RC(t,\theta) = \frac{1}{2} \frac{EI(t+\Delta t,\theta) - EI(t,\theta)}{EI(t+\Delta t,\theta) + EI(t, \theta)},
    \label{eqn:rc}
\end{equation}

where $\Delta t$ is the time step, and $EI(t,\theta)$ is the elastic impedance at time $t$ and incident angle $\theta$. EI in equation \ref{eqn:rc} refers to the normalized elastic impedance proposed by \cite{whitcombe2002elastic} which is computed from the elastic properties as follows:  
\begin{equation}
    EI(t, \theta) = V_{p_0} \rho_0 \left(\frac{V_p(t)}{V_{p_0}}\right)^a\left(\frac{V_s(t)}{V_{s_0}}\right)^b\left(\frac{\rho(t)}{\rho_0}\right)^c,
\end{equation}
where, 
\begin{minipage}{0.5\linewidth}
\begin{align*}
a &= {1+\tan^2\theta}\\
b &= {4K\sin^2\theta}\\
\end{align*}
\end{minipage} 
\begin{minipage}{0.5\linewidth}
\begin{align*}
c &= {1-4K\sin^2\theta}\\
K &= \nicefrac{V_s^2}{V_p^2}\\  
\end{align*}
\end{minipage} 
and 
$V_p$,$V_s$ and $\rho$ are P-wave velocity, S-wave velocity, and density, respectively, and $V_{p_0}$,$V_{s_0}$ and $\rho_0$ are their respective averages over the training sample (i.e., well-logs). It is worth noting that the Aki-Richard's approximation of the elastic impedance is only valid for incident angles that are less than $35^\circ$. Thus, in the case study, we only consider valid angles for this approximation.  

The seismograms are then generated by convolving $RC(t,\theta)$ with a wavelet $w(t)$, i.e., 
\begin{equation}
    S(t, \theta) = w(t) * RC(t, \theta), 
    \label{eqn:seismic}
\end{equation}

where $*$ is the linear convolution operator, and $S(t,\theta)$ is the synthesized seismogram. Thus, the forward model utilized in this work is a 1-dimensional convolutional forward model that synthesized seismograms from EI traces. 

\section{Case Study on Marmousi 2 Model}
In order to validate the proposed algorithm, we chose Marmousi 2 model as a case study. Marmousi 2 model is an elastic extension of the original Marmousi synthetic model that has been used for numerous studies in geophysics for various applications including seismic inversion, seismic modeling, seismic imaging, and AVO analysis. The model spans 17 km in width and 3.5 km in depth with a vertical resolution of 1.25 m. The details of the model can be found in \cite{martin2006marmousi2}. 

In this work, we used the elastic model (converted to time) to generate EI and seismic data to train the model. The details of the dataset generation are discussed in the next section. The proposed workflow is trained using all seismic traces in Marmousi 2 in addition to a few EI traces (training traces). The workflow is then used to invert for EI on the entire Marmousi. Since the full elastic model is available, we compare our EI inversion with the true EI quantitatively.

\subsection{Dataset Generation}
We used the elastic model of Marmousi 2 to generate EI for 4 incident angles $\theta=0^{\circ}$,$10^{\circ}$,$20^{\circ}$, and $30^{\circ}$. Thus, in Figure \ref{fig:inverse}, $c_1 =4$ which represents the number of channels in the input traces. Multi-angle seismic data (a total of $N_s=2720$ traces) is then generated from EI using the forward model with Ormsby wavelet (5-10-60-80 Hz) following the synthesis procedure in \cite[]{martin2006marmousi2}. The seismic traces are then downsampled (in time) by a factor of 6 to simulate resolution difference between seismic and well-log data. Finally, a $15$ db white Gaussian noise is added to assess the robustness of the proposed workflow to noise. 

\begin{figure}[ht]
    \centering
    \includegraphics[width=\linewidth]{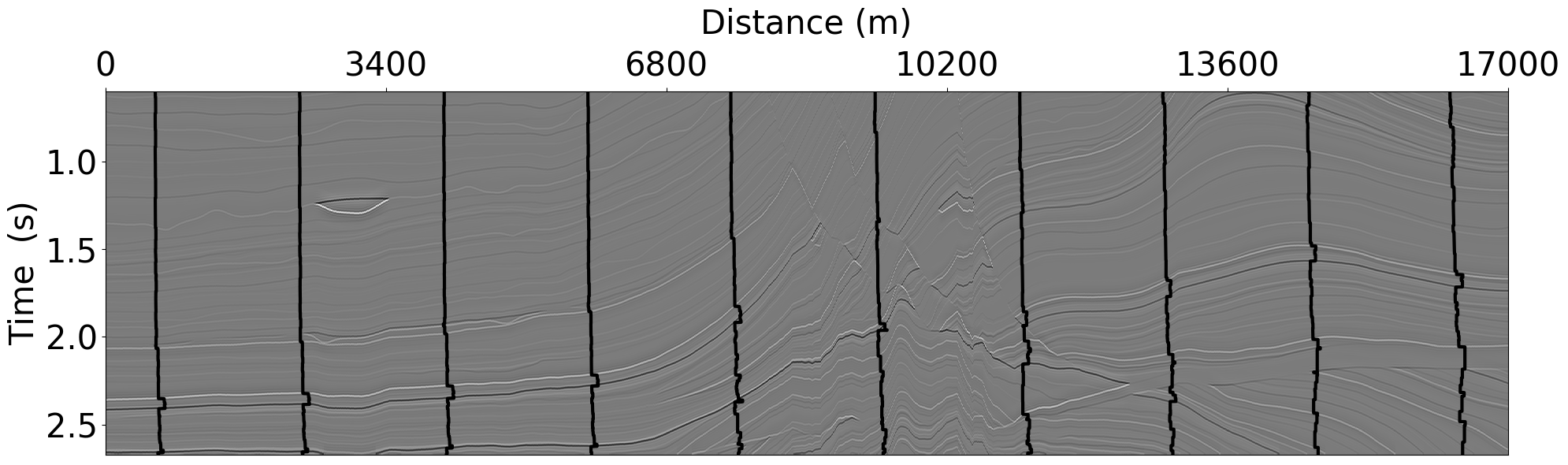}
    \caption{The training EI traces overlaid on the zero-offset seismic section.}
    \label{fig:well_locations}
\end{figure}

To train the proposed inversion workflow, we chose $10$ evenly-spaced traces for training ($N_p=10$) as shown in Figure \ref{fig:well_locations}. We assume we have access to both EI and seismic data for the training traces. For all other traces in the survey, we assume we have access to seismic data only. Although using more EI traces for training would improve the inversion, in this work we use only 10 EI traces for training to simulate well-log data in a practical setting. 

\subsection{Training The Inverse Model}
 First, the inverse model (neural network) is initialized with random parameters. Then, randomly chosen seismic traces in addition to the seismic traces for which we have EI traces in the training dataset are inputted to the inverse model to get a corresponding set of EI traces. The forward model is then used to synthesize seismograms from the estimated EI. \emph{Seismic loss} is computed as the MSE between the synthesized seismic and the input seismic. Moreover, \emph{property loss} is computed as the MSE between the predicted EI and the true EI trace on the training traces only. The total loss is computed as the sum of the two losses. Then, the gradients of the total loss are computed, and the parameters of the inverse model are updated accordingly. The process is repeated until convergence. 

Figure \ref{fig:inverse} shows the inverse model with no given hyperparameters to ensure its generalizability for data other than the data used in this case study. These hyperparameters are \{$c_1,c_2, s_1, s_2, d_1, d_2$,$k$\} which are parameters of the inversion model that need to be set rather than learned. Some of these hyperparameters are completely dictated by the data. For example, $c_1$ which is the number of inputs channels must be chosen as the number of incident angles of the data. Also, $s_1, s_2$ must be chosen such that $s_1\times s_2$ is equal to the resolution mismatch factor between seismic and EI data. In this case study, we set $c_1=4$, $s_1=3$, and $s_2=2$. The rest of the hyperparameters are chosen by analyzing the data and trying different values and testing the performance on a validation dataset using cross-validation. In this case study, we choose $c_2=8$, $k=5$, $d_1=1$, $d_2=3$, and $d_3=6$.

\subsection{Results and Discussion}
\begin{figure*}[ht!]
    \centering
    \input{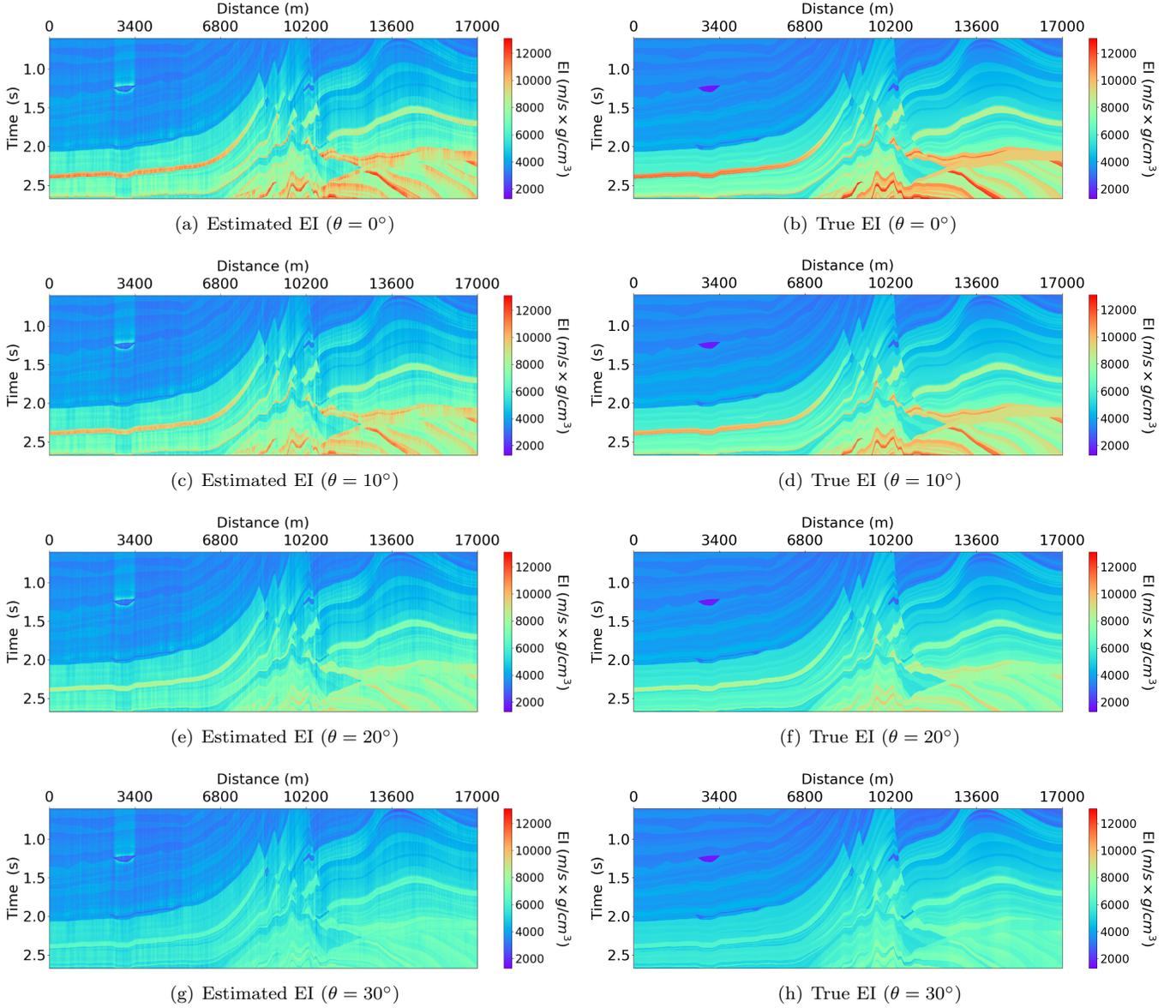}
    \caption{Estimated EI and true EI for Marmousi 2 model.}
    \label{fig:results_section_EI}
\end{figure*}

\begin{figure*}[ht!]
    \centering
    \input{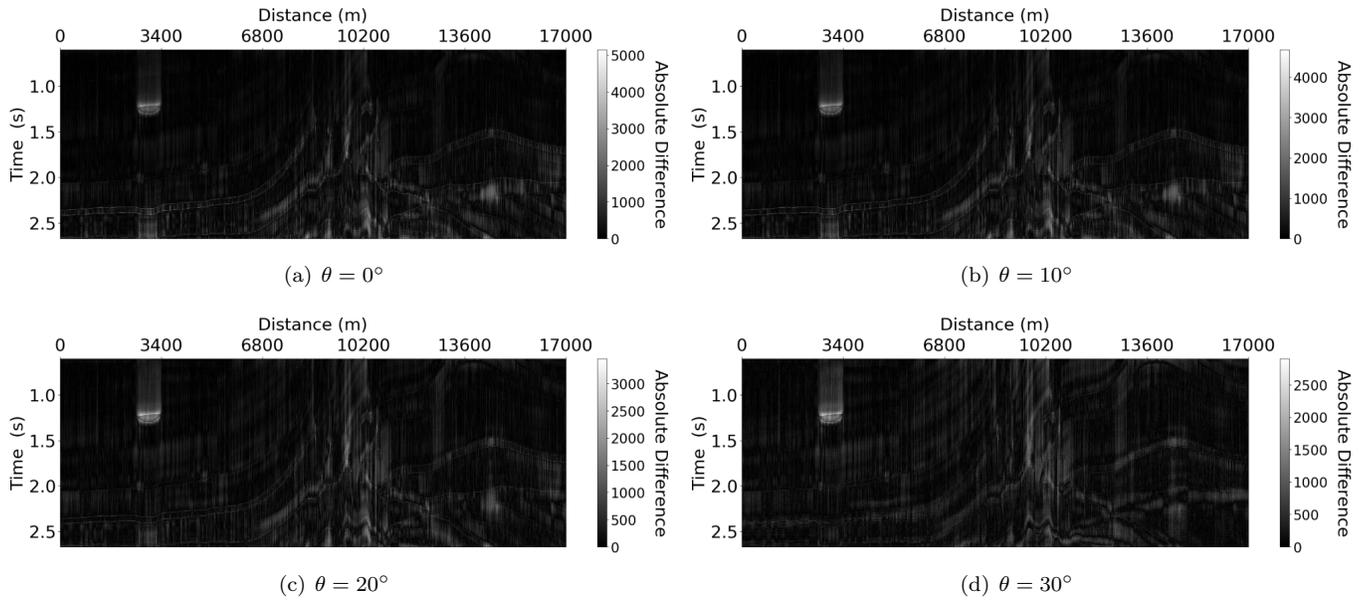}
    \caption{Absolute difference between True EI and estimate EI for all incident angles.}
    \label{fig:results_diff_EI}
\end{figure*}

Figure \ref{fig:results_section_EI} shows estimated EI and true EI for the entire section for all four incident angles. Figure \ref{fig:results_diff_EI} shows the absolute difference between the true and estimated EI. The figures indicate that the proposed workflow estimates EI accurately for most parts of the section with a visible lateral jitter. Jitter effect is expected since the proposed workflow is based on 1-dimensional sequence modeling with no explicit spatial constraints as often done in classical inversion methods. Furthermore, the noise in the seismic data reduces the similarity between neighboring seismic traces, which can also cause such jitter. The shown sections are the direct output of the inversion workflow with no post-processing which can reduce such artifacts.

\begin{figure*}[ht!]
    \centering
    \input{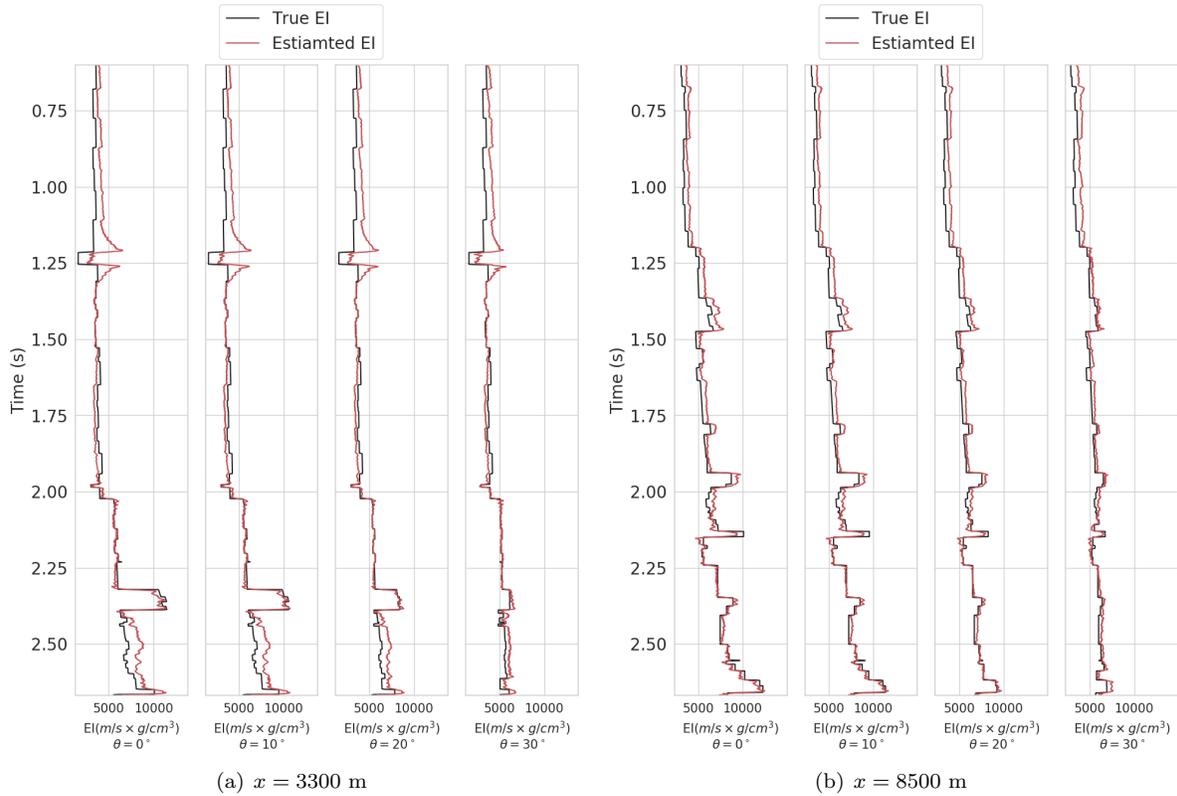}    
    \caption{Selected EI trace. Estimate EI is shown in red, and true EI is shown in black. }
    \label{fig:results_traces}
\end{figure*}

Figure \ref{fig:results_traces} shows two selected traces from the section that were not in the training dataset ($x=3200$ m and $x=8500$ m). The trace at $x=3200$ passes through an anomaly (Gas-charged sand channel) represented by an isolated and sudden transition in EI at $1.25$ seconds. This anomaly causes the inverse model to incorrectly estimate EI. Since our workflow is based on bidirectional sequence modeling, we expect the error to propagate to nearby samples in both directions. However, the algorithm quickly recovers a good estimate for deeper and shallower samples of the trace. This quick recovery is mainly due to the reset-gate variable in the GRU that limits the propagation of such errors in sequential data estimation. Furthermore, the trace at $x=8500$ passes through most layers in the section which makes it a challenging trace to invert. The estimated EI at $x=8500$ follows the trends in true EI trace, including the thin layers.

\begin{figure}[ht!]
  \centering
    \input{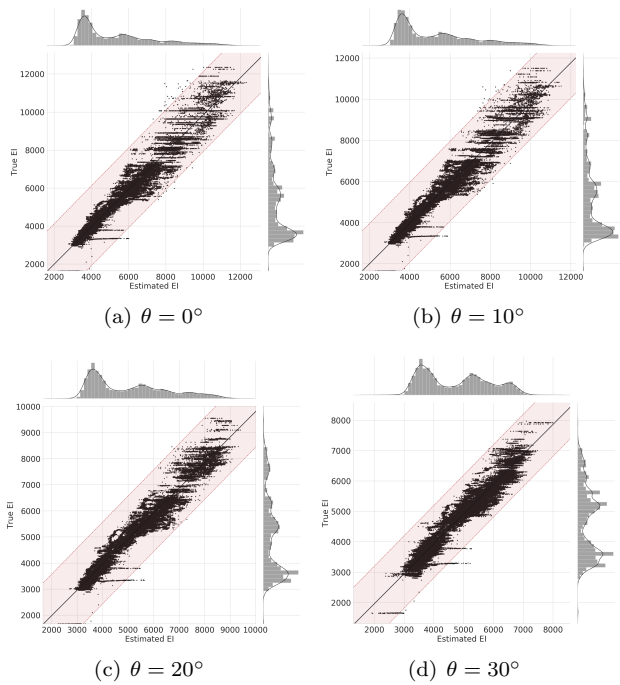}
    \caption{Scatter plots of the estimated and true EI for different values of $\theta$. The shaded regions include all points that are within $\pm \sigma_{\text{EI}}$ of the true EI.}
    \label{fig:results_correlation}
\end{figure}

Furthermore, we show scatter plots of the estimated and true EI for all incident angles in Figure \ref{fig:results_correlation}. The shaded regions in the figure include all points that are within one standard deviation of the true EI ($\sigma_{\text{EI}}$). The scatter plots show a linear correlation between the estimated and true EI with the majority of the estimated samples within $\pm\sigma_{\text{EI}}$ from the true EI.

To evaluate the performance of the proposed workflow quantitatively, we use two metrics that are commonly used for regression analysis. Namely, Pearson correlation coefficient (PCC), and coefficient of determination ($r^2$). PCC is defined as: 
\begin{equation}
    \text{PCC} =\frac{1}{T}\frac{1}{\sigma_x\sigma_{\hat{x}}}\sum_{t=0}^{T-1} \left[x(t)-\mu_x\right]\left[\hat{x}(t)-\mu_{\hat{x}}\right], 
\end{equation}

where $x$ and $\hat{x}$ are the target trace and the estimated trace, respectively, and $\mu$ and $\sigma$ are mean and standard deviation, respectively. PCC is a measure of the linear correlation between the estimated and target traces. It is commonly used to measure the overall fit between the two traces with no regard to the individual values. On the other hand, $r^2$ is a goodness-of-fit measure, where it takes into account the mean squared error between the two traces. The coefficient of determination ($r^2$) is defined as: 

\begin{equation}
    r^2 = 1-  \frac{\sum_{i=0}^{T-1}\left[x(t)-\hat{x}(t)\right]^2}{\sum_{i=0}^{T-1}\left[x(t)-\mu_x\right]^2}
\end{equation}

In addition, we used Structural Similarity (SSIM) \cite[]{wang2004image} to assess the quality of the estimated EI section from an image point of view. SSIM evaluates the similarity of two images from local statistics (on local windows) using the following equation:
\begin{equation}
    \text{SSIM}(\mathbf{X},\mathbf{\hat{X}}) = \left[l(\mathbf{x},\mathbf{\hat{x}})\right]^\alpha \cdot \left[c(\mathbf{x},\mathbf{\hat{x}})\right]^\beta \cdot
    \left[s(\mathbf{x},\mathbf{\hat{x}})\right]^\gamma, 
    \label{eqn:ssim}
\end{equation}

where $\mathbf{x}$ and $\mathbf{\hat{x}}$ are patches from estimated and target image, respectively. $l(\mathbf{x},\mathbf{y})$, $c(\mathbf{x},\mathbf{\hat{x}})$, and $s(\mathbf{x},\mathbf{\hat{x}})$ are luminance, contrast and structure comparison functions respectively. $\alpha>0$, $\beta>0$, $\gamma>0$ are constants chosen to adjust the influence of the each of three terms. From SSIM, a single similarity value, denoted as M-SSIM, can be computed by taking the mean of all SSIM scores for all local windows. 

The quantitative results are summarized in Table \ref{tab:results}. It is evident that the estimated EI captures the overall trend of true EI (average PCC $\approx 98\%$, average $r^2\approx94\%$). The average $r^2$ score is lower than PCC since it is more sensitive to the residual sum of squares rather than the overall trend. Note that PCC and $r^2$ compute scores over individual traces, then the final score (for each incident angle) is computed by averaging across all traces. On the other hand, M-SSIM is computed over the 2D section (image) which takes both lateral and vertical variations into account.

\begin{table}[ht!]
    \centering
    \caption{Quantitative evaluation of the proposed workflow on Marmousi 2.}
    \label{tab:results}
    \begin{tabular}{c|c|c|c}
    \hline
    \hline
    \diagbox{\textbf{Angle}}{\textbf{Metric}}& \textbf{PCC} & $\boldsymbol{r}^2$ & \textbf{M-SSIM} \\
    \hline
    $\theta=0^{\circ}$  &0.98 &0.95 &0.92\\
    $\theta=10^{\circ}$ &0.98 &0.95 &0.92\\
    $\theta=20^{\circ}$ &0.98 &0.95 &0.92\\
    $\theta=30^{\circ}$ &0.98 &0.92 &0.92\\
    \hdashline
    Average             &0.98 &0.94 &0.92\\
    \hline
    \hline
\end{tabular}

%EI
% \begin{tabular}{c|c|ccc}
%     \hline
%     \hline
%     &\diagbox{\textbf{Angle}}{\textbf{Metric}}& \textbf{PCC} & $\boldsymbol{r}^2$ & \textbf{M-SSIM} \\
%     \hline
%     \multirow{5}{*}{\rotatebox[origin=c]{90}{\footnotesize{Entire Section}}}
%     &$\theta=0^{\circ}$  &0.9811 &0.9477 &0.9157\\
%     &$\theta=10^{\circ}$ &0.9821 &0.9487 &0.9165\\
%     &$\theta=20^{\circ}$ &0.9834 &0.9459 &0.9153\\
%     &$\theta=30^{\circ}$ &0.9776 &0.9199 &0.9194\\
%     \cline{2-5}
%     &Average             &0.9811 &0.9406 &0.9167\\
%     \cmidrule{2-5}\morecmidrules\cmidrule{2-5}
%     \multirow{5}{*}{\rotatebox[origin=c]{90}{\footnotesize{Training Traces}}}
%     &$\theta=0^{\circ}$  &0.9951 &0.9900 &-\\
%     &$\theta=10^{\circ}$ &0.9955 &0.9909 &-\\
%     &$\theta=20^{\circ}$ &0.9955 &0.9904 &-\\
%     &$\theta=30^{\circ}$ &0.9896 &0.9763 &-\\
%     \cline{2-5}
% %    \hdashline
%     &Average             &0.9940 &0.9870 &-\\
%     \hline
%     \hline
% \end{tabular}

%%Seismic
% \begin{tabular}{|c|c|c|c|}
%     \cline{1-4}
%     \diagbox{Angle}{Metric}& PCC & $r^2$ & M-SSIM \\
%     \hline
%     $\theta=0^{\circ}$  &0.9888 &0.9773 &0.9941\\
%     $\theta=10^{\circ}$ &0.9881 &0.9759 &0.9943\\
%     $\theta=20^{\circ}$ &0.9826 &0.9650 &0.9934\\
%     $\theta=30^{\circ}$ &0.9889 &0.9771 &0.9976\\
%     Average             &0.9871 &0.9738 &0.9948\\
%     \hline
% \end{tabular}

\end{table}

Furthermore, we quantify the contribution of each of the two terms in the loss function in equation \eqref{eqn:semi-supervised} by comparing the inversion results for different values of $\alpha$ and $\beta$. The average results are reported in Table \ref{tab:semisupervised_vs_supervised}.

\begin{table*}[ht]
\centering
\caption{Quantitative evaluation of the sensitivity of the loss function with respect to its two terms. \textbf{Unsupervised}: learning from all seismic traces in the section, \textbf{Supervised}: learning from 10 seismic traces and their corresponding EI traces, \textbf{Semi-supervised}: learning from all seismic data in the section in addition to 10 seismic traces and their corresponding EI traces}
\label{tab:semisupervised_vs_supervised}
\begin{tabular}{c|c|c|c}
    \hline
    \hline
    \diagbox{\textbf{Training Scheme}}{\textbf{Metric}}& \textbf{PCC} & $\boldsymbol{r}^2$ & \textbf{M-SSIM}\\
    \hline
    Unsupervised ($\alpha=0, ~\beta=1$) & 0.33 & -0.45 & 0.77\\
    Supervised ($\alpha=1,~\beta=0$) & 0.96 & 0.88 & 0.87\\
    Semi-supervised ($\alpha=1, ~\beta=1$) &0.98 & 0.94 & 0.92\\
    \hline
    \hline
\end{tabular}
\end{table*}

When $\alpha=0$ and $\beta=1$ (unsupervised learning), the inverse model is learned completely from the seismic data without integrating the well-logs. Thus, the results are the worst out of three schemes. Note that the performance of the unsupervised learning scheme could be improved by using an initial smooth model and by enforcing some constraints on the inversion as often doe in classical inversion. Moreover, when $\alpha=1$ and $\beta=0$ (supervised learning), the inversion workflow learns from only 10 seismic traces and their corresponding EI traces from well-logs. Thus, it is expected that it results in better inversion compared to the unsupervised scheme. However, training a deep inverse model in a supervised learning scheme requires heavy regularization and careful selection of the training parameters. In addition, the learned inverse model might not generalize well beyond training data. Finally, when $\alpha=\beta=1$ (semi-supervised learning), the inverse model is learned from all seismic data in addition to 10 EI traces from well-logs. Hence, the semi-supervised learning scheme improves the performance and regularizes the learning. 

Figure \ref{fig:dist_corr} shows the distribution of PCC of the estimated EI with respect to the true EI for all traces in Marmousi 2 model. The figure shows that the estimated EI correlates very well with the true EI, with the majority of PCC values between $[0.9 - 1]$, and a spike near $1$. Similarity, Figure \ref{fig:dist_r2} shows the distribution of $r^2$ values, with a wider distribution than that of PCC. This is mainly due to the fact that $PCC$ is defined to be in the range $[0,1]$ and $r^2$ is in the range $(-\infty, 1]$. In addition, $r^2$ is a more strict metric than PCC as it factors in the MSE between the estimated EI and true EI. Figure \ref{fig:dist_ssim} shows the distribution of local SSIM scores over the entire section, with the majority of the scores in the range $[0.8,1]$ indicating that the estimated EI is structurally similar to the true EI from an image point of view. 

\begin{figure}[ht!]
  \centering
    \input{Fig/EI_distribution.tex}
    \caption{The distribution of Pearson correlation coefficient (PCC), and coefficient of determination ($r^2$), and SSIM values on Marmousi 2. }
    \label{fig:results_distribution}
\end{figure}

\subsection{Implementation} 
\begin{figure}[ht!]
  \centering
    \includegraphics[width=\linewidth]{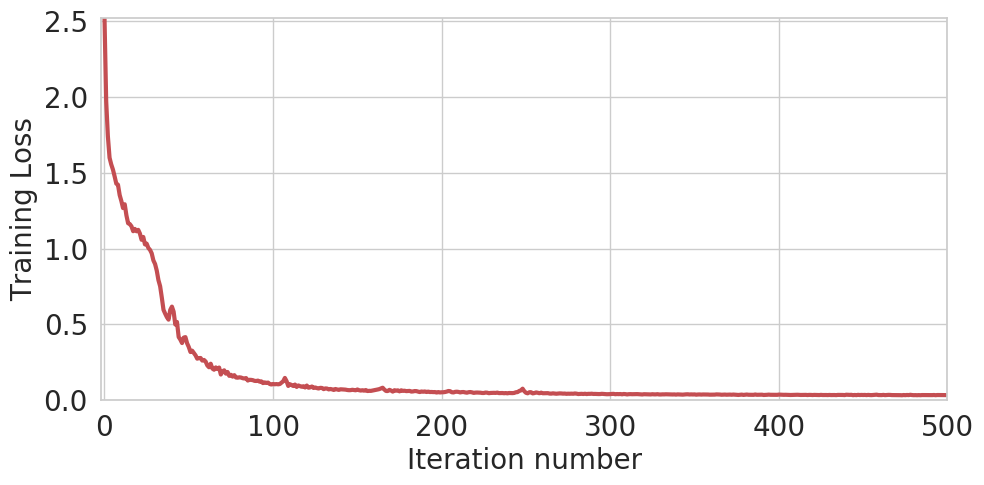}
    \caption{Training learning curve showing the loss function value over 500 iterations.}
    \label{fig:learning_curve}
\end{figure}

The proposed workflow was implemented in Python using PyTorch deep learning library \cite[]{paszke2017automatic}. For optimization, we used Adam \cite[]{kingma2014adam} which is a gradient-based stochastic optimization technique with adaptive learning rate that was designed specifically for training deep neural networks. The codes were run on a PC with Intel i7 Quad-Core CPU, and a single Nvidia GeForce GTX 1080 Ti GPU. The run time of 500 iterations of the GPU-accelerated algorithm was 2 minutes. Figure \ref{fig:learning_curve} shows the convergence behavior of the proposed workflow over 500 training iterations. The y-axis shows the total loss over the training dataset which is computed as in equation \ref{eqn:loss} except for an additional normalization factor that is equal to the number of time samples per trace. Note that the loss is also computed over normalized traces after subtracting the mean value and dividing by the standard deviation to ensure faster convergence of the inversion workflow as often done for deep learning models. The proposed workflow converges in about 300 iterations. However, the loss continues to decrease slightly. The training was terminated at after 500 iterations to avoid over-fitting to the training dataset. 

The efficiency of the proposed workflow is due to the use of 1-dimensional modeling. In addition, Marmousi 2 is a relatively small model with 2720 traces only. However, the computation time of the proposed workflow will scale linearly with the number of traces in the dataset. 

The code used to reproduce the results reported in this manuscript is publicly available on GitHub \href{https://github.com/olivesgatech/Elastic-Impedance-Inversion-Using-Recurrent-Neural-Networks}{[\color{blue}{\underline{link to code}}]}.  

\section{Conclusions}
In this work, we proposed an innovative semi-supervised machine learning workflow for elastic impedance inversion from multi-angle seismic data using recurrent neural networks. The proposed workflow was validated on the Marmousi 2 model. Although the training was carried out on a small number of EI traces for training, the proposed workflow was able to estimate EI with an average correlation of $98\%$. Furthermore, the applications of the proposed workflow are not limited to EI inversion; it can be easily extended to perform full elastic inversion as well as property estimation for reservoir characterization.

\section{Acknowledgment}
This work is supported by the Center for Energy and Geo Processing (CeGP) at Georgia Institute of Technology and King Fahd University of Petroleum and Minerals (KFUPM).

% \append{}
% \input{5.Appendix.tex}

%\verbatiminput{references.bib}

\bibliographystyle{seg}  % style file is seg.bst
\bibliography{references}

\end{document}